%% file: 4.arxiv.tex
\documentclass{tlp}%
\pdfoutput=1

\renewcommand\cite[1]{\citet{#1}}

\newcommand{\myenumeratefix}{
  \setlength\leftmargin{25pt}%
}
\usepackage{mytlpnojournal4}

\usepackage{xspace}
\usepackage{latexsym}
\usepackage{amssymb}   
\usepackage{abraces}   

\usepackage{color}

\newcommand{\grey}{\color[rgb]{.7, .7, .7}}
\newcommand{\mypalegrey}{\color[rgb]{.9, .9, .9}}
\newcommand\black {\color{black}}

\usepackage{hyperref}

\newtheorem{theorem}{Theorem}
\newtheorem{example}[theorem]{Example}
\newtheorem{definition}[theorem]{Definition}
\newtheorem{lemma}[theorem]{Lemma}
\newtheorem{proposition}[theorem]{Proposition}

\newcommand{\myparagraph}[1]{%
  \paragraph{#1}}

\newcommand*{\seq}[2][n]  {{#2_{1}, \allowbreak \ldots, \allowbreak #2_{#1}}}

\newcommand*{\notmodels}{\mathrel{\,\not\!\models}}

\newcommand*{\myprologneg}{\ensuremath{\mbox{\tt \symbol{92}+}}\xspace}
\newcommand*{\myunderscore}{\mbox{\tt\symbol{95}}}

\newcommand*{\HU}{{\ensuremath{\cal{H U}}}\xspace}
\newcommand*{\HB}{{\ensuremath{\cal{H B}}}\xspace}

\newcommand*{\M}{{\ensuremath{\cal M}}\xspace}

\newcommand{\ttt}{\ensuremath{{\bf t}}\xspace}
\newcommand{\fff}{\ensuremath{{\bf f}}\xspace}
\newcommand{\tttfff}{\ensuremath{{\bf tf}}\xspace}
\newcommand{\uuu}{\ensuremath{{\bf u}}\xspace}

\newcommand*{\st}{{\ensuremath{\it St}}\xspace}
\newcommand*{\snf}{{\ensuremath{\it Snf}}\xspace}
\newcommand*{\snfa}{{\ensuremath{\it Snf}\hspace{-2pt}}\xspace}

\newcommand*{\NN}{{\ensuremath{\mathbb{N}}}\xspace}

\begin{document}

\jnlPage{1}{17}
\jnlDoiYr{2025}

\newcommand{\Version}{Version 1.12}

\title [Systematic construction of programs]
{On systematic construction\\ of correct logic programs}
\lefttitle{W. Drabent}

\begin{authgrp}
\author%
       {W{\l}odzimierz Drabent\\\mbox{}%
}
\affiliation{%
  Institute of Computer Science,    Polish Academy of Sciences
}

\end{authgrp}

%
%

\history{2025-07-27}

\maketitle

\begin{abstract}

Partial correctness of imperative or functional programming divides in logic
programming into two notions.
Correctness means that all answers of the program are compatible with the
specification.  Completeness means that the program produces all the answers
required by the specifications.
We also consider semi-completeness -- completeness for those queries for
which the program does not diverge.
This paper presents an approach to
systematically construct provably correct and semi-complete logic programs,
for a given specification.  Normal programs are considered, under
Kunen's 3-valued completion semantics (of negation as finite failure) and
the well-founded semantics (of negation as possibly infinite failure).
The approach is declarative, it abstracts from details of operational
semantics, like e.g.\ the form of the selected literals (``procedure calls'')
during the computation.
The proposed method is simple, and can be used (maybe informally) in actual
everyday programming.

\end{abstract}


\medskip\noindent
This paper presents a method of constructing provably correct and
semi-complete normal logic programs.  The considered semantics are Kunen
semantics (3-valued completion semantics, corresponding to negation as finite
failure, NAFF), and the well-founded semantics (corresponding to negation as,
possibly infinite, failure).
The approach is declarative, it abstracts from details of operational
semantics, like e.g.\ the form of the selected literals (``procedure calls'')
during the computation.  The approach can be used, possibly at an informal
level, in programmers' practice.
It can be seen as a guidance on how to construct programs, and thus
used in teaching logic programming.

In this paper, we first   %
recall the method for definite clause programs, presented earlier.
Then ``Preliminaries'' present the main concepts, including both semantics of
interest.  The next section discusses specifications for programs with
negation.
Then, for each semantics, we present ways of proving correctness and
completeness, and a method of constructing programs.
We assume that the reader is familiar with basics of logic programming,
negation in logic programming, and Prolog.

\section{Constructing definite clause programs}
\label{sec:definite}
Here we briefly present a method of \cite{drabent.tplp18}.
For this we first
 informally describe the main concepts related to program
correctness.  For further details and references see
the papers by \cite{drabent.tocl16,drabent.tplp18} and
\cite{Apt-Prolog-short}). 
 Definite clause programs will be called shortly {\em positive programs}.

In imperative (or functional) programming, a central notion is partial
correctness.  A program is partially correct if its termination implies its
compatibility with the specification.
In logic programming this divides into two notions: correctness and
completeness.  
  Correctness of a program $P$ w.r.t.\ (with respect to) a specification $S$  means that each
  answer of $P$ is compatible with $S$,
  completeness -- that each answer required by $S$ is an answer of $P$.
More technically:
A specification is an Herbrand interpretation.  
A program $P$ with the least
Herbrand model $\M_P$ is {\em correct}
w.r.t.\ a specification $S$ if $\M_P\subseteq S$.
It is {\em complete} w.r.t.\ $S$ if $S\subseteq\M_P$.  

By an answer of a positive
program $P$ we mean any query $Q$ such that $P\models Q$.
(Answer is sometimes called correct, or computed, instance of a query
\citep[Def.\,4.6,\,3.6]{Apt-Prolog-short}.)
In practice, if Prolog succeeds for $P$ and a query $Q_0$ with a substitution
$\theta$ then $Q_0\theta$ is an answer for $P$.
For a program $P$ correct w.r.t.\,$S$, each answer $Q$ of $P$ is
compatible with $S$
 (i.e.\ $S\models Q$); 
for $P$  complete w.r.t.\,$S$,
each ground answer $Q$ required by $S$ (i.e.\ $S\models Q$) is an answer of $P$.

\myparagraph{\bf Proving correctness.}
There exists a simple (however often forgotten) way of proving correctness of
positive programs: 
A sufficient condition for 
$P$ being correct w.r.t.\ $S$ is $S\models P$ \citep {Clark79}.
(In other words, the sufficient condition is that for each ground instance
of a clause of $P$, if the body atoms are in $S$ then the head is in $S$.)
Informally: any clause applied to correct atoms produces a correct atom.

\myparagraph{\bf Proving completeness.}
We need some %
auxiliary notions.  
A ground atom $A$ is {\em covered} \citep{Shapiro.book} w.r.t.\ $S$
by a clause $C$, if $C$ has a ground instance
 $A\mathop\gets\seq B$ and $\seq B\in S$ %
(informally: if $A$ can be produced by $C$ from $S$).
Given a specification $S$, program
$P$ is {\em complete for a query} $Q$ %
if, speaking informally, it produces all the answers for $Q$ required by $S$.  
$P$ is  {\em semi-complete} if it is complete
for any query for which there exists a finite SLD-tree.
To reason about completeness, it is convenient %
to deal with semi-completeness and termination separately.
Often a proof of completeness is similar to a proof of semi-completeness
together with a proof of termination; 
and the termination has to be established anyway.

Again, there is a simple sufficient condition for semi-completeness:
If each atom from $S$ is covered %
w.r.t.\ $S$ by
 a clause of $P$ then
$P$ is semi-complete w.r.t.\ $S$ \citep{drabent.tocl16}.
Informally: any atom required to be produced can be produced
by a clause of the program out of atoms required to be produced.

It is surprising that reasoning about completeness was often
neglected.  E.g.\ an important monograph by \cite{Apt-Prolog-short} does not
even 
mention the notion of program completeness.
Completeness for %
positive programs was dealt with by
\cite
{Deransart.Maluszynski93},
\cite{DBLP:journals/tplp/DrabentM05shorter}, \cite{drabent.tocl16},
and for normal programs by \cite{Malfon} and
\cite{DBLP:journals/tplp/DrabentM05shorter}.
\cite{Sterling-Shapiro-shorter} discuss
 completeness informally.

\myparagraph{\bf Approximate specifications.}
An important observation is that many issues can be simplified if we consider 
separate specifications for correctness and completeness
\citep{DBLP:journals/tplp/DrabentM05shorter,drabent.tocl16,drabent.tplp18}.
A single specification for both has to exactly describe the least Herbrand
model of a 
program.  This is often too troublesome (and not necessary).
For instance, it is unimportant how {\it append}/3 of Prolog behaves on
non list arguments,
or what happens when the second argument of {\it member}/2 is not a list,
 or how {\it insert}/3 of insertion sort inserts a number
into a not sorted list.
Moreover, in program construction
we should not fix in advance the behaviour of the future program in such
unimportant cases. (In the main example of the paper by
\cite{drabent.tplp18}, such 
behaviour is different in consecutive versions of the program; otherwise
an efficient version could not have been obtained.)
A pair 
 $(S_{\it c o m p l},S_{corr})$ 
of specifications for completeness and correctness, where 
$S_{\it c o m p l}\subseteq S_{corr}$, will be called 
{\em approximate specification}.
A program with the least Herbrand model $\M_P$ will be called
{\em fully correct} w.r.t.\ $(S_{\it c o m p l}, S_{corr})$ if
$S_{\it c o m p l}\subseteq\M_P\subseteq S_{corr}$.

\begin{example}
\label{ex.approx}
  A list membership predicate $m$/2 should define the set
  $r= \{\, m(e_i,[\seq e])\in\HB \mid\linebreak[3] 1\leq i\leq n \,\}$
  (\HB is the Herbrand base).
  However defining exactly $r$ would be inefficient, and the {\it member}/2
  of Prolog
  defines a superset of $r$.  It is not important what exactly the defined set
  is. 
  What matters is that whenever the second argument is a list, the first
  argument is its member.
  Hence any ground answer should be a member of
  $r_+ =\linebreak[3]
  \{\, m(e,u)\in\HB \mid \mbox{if $u$ is a list}\ \mbox{then}\
  \mbox{$m(e,u)\in r$} \,\}
  \linebreak[3]
  = r\cup \{\, m(e,u)\in\HB \mid u \mbox{ is not a list} \,\}
  $.
  This leads to an approximate specification %
  $(r,r_+)$.  It states that the atoms from $r$ (resp.\,$r_+$)
  must (may) be answers of the program.
\end{example}

\myparagraph{\bf Program construction.}
The above sufficient conditions for semi-completeness and correctness suggest
a method \citep{drabent.tplp18} of constructing a program, given a specification 
$(S_{\it c o m p l},\linebreak[3] S_{corr})$.  
Due to the condition for semi completeness, we need (A) to construct clauses
$\seq C$ so that each atom from $S_{\it c o m p l}$ is covered 
w.r.t.\ $S_{\it c o m p l}$ by some of them.
For correctness we need that (B) $S_{corr}\models C_i$ for each such $C_i$.
This leads to a provably correct and semi-complete program.  
We also need that (C) the clauses are chosen so that the program terminates 
for queries of interest.

Termination is kept outside of the proposed rigorous method.  One may use any
way of proving termination; when applying the method informally, one uses
the usual hints of Prolog programming craft, like using in the body proper
subterms of the terms in the head.

\begin{example}
\label{ex.member}
Take specification $(r,r_+)$ from Ex.\ \ref{ex.approx}.
  For (A) consider an atom $A\in r$.  We have two cases.
1.\,$A=m(e,[e|t])$.  Such $A$ is covered by a fact
$C_1 = m(E,[E|\myunderscore])$;
note that (B) holds.  2.\,$A=m(e,[e'|t])$, and $e$ is a member of the list $t$.
This suggests a clause $C_2=\linebreak[3] m(E,[\myunderscore|T])\mathop\gets m(E,T)$.  
It covers $A$, and (B) holds.  As  $T$ is a proper subterm of $[\myunderscore|T]$, 
program $M=\{C_1,C_2\}$ terminates for any ground query. 
Thus $M$ is 
correct and complete w.r.t.\  $(r,r_+)$.
Example wrong choices in step 2.\ are: a fact $m(E,[F|T])$ (covers $A$, but
(B) does not hold), a clause
$m(E,[\myunderscore|T])\mathop\gets m(E,[\myunderscore|T])$ (fulfils (A)
and
(B), but leads to non-termination).
Note that we constructed the built-in procedure {\it member}/2 of Prolog.
\end{example}

\begin{example}
  For a bigger and detailed example, consider finding the middle element of a
  list, without using numbers in the program.
  The specification is  $(S_{mid},S_{mid})$ where 
  $S_{mid}=\{\,mid(e_i,[\seq[2i-1]e])\in\HB \mid i>0 \,\}$.
We follow a standard hint -- solve
a more general problem.
Instead of %
a list of a certain length containing $e$ at a certain position,
let us
consider a term with $e$ at this position, and a list of this length.
This leads to a specification of a new predicate: $(S_{mi},S_{mi})$, where
$
S_{mi}=\{\,mi(e_i,[\seq[i]e|t],[\seq[2i-1]f])\in\HB\mid i>0\,\}
$, and specification $(S,S)$,  where $S=S_{mid}\cup S_{mi}$, for the program.
For (A) we have three cases:

1.\;$A=mid(e_i,l)$, where $l=[\seq[2i-1]e]$.  Obviously
$mi(e_i,l,l)\in S$. So
$A$ is covered by $C_1 = mid(E,L)\gets mi(E,L,L)$.
To show (B), note that if $mi(e,l',l')\in S$ then $l'=[\seq[2i-1]f]$ and $e=f_i$,
hence $mid(e,l')\in S$.  Thus $S\models C_1$.

2.\;$A=mi(e,[e|t],[f])$.
Covered by $C_2= mi(E,[E|\myunderscore],[\myunderscore])$.
Note that $S\models C_2$.

3.\;$A=mi(e_i,[\seq[i]e|t],[\seq[2i-1]f])$, where $i>1$.
For (A) and (C) we
should employ atom(s) from $S$, preferably with arguments which are
subterms of those of $A$.   A candidate argument
 is $u=[e_2,\ldots,e_i|t]$.
Note that $mi(e_i,u,l)\in S$ whenever $l$ is a list of length $2i-3$.
Hence  $A$ is covered
by
$C_3=mi(E,[\myunderscore|U],[\myunderscore,\myunderscore|L]) \gets
mi(E,U,L)$.
For (B) note that if  $mi(e,u',l)\in S$ then each term
$mi(e,[t_1|u'],[t_2,t_3|l])$ is in $S$.  Hence $S\models C_3$.

The obtained program is $\{C_1,C_2,C_3\}$. 
As it terminates for ground queries (we skip an easy proof), 
it is correct and complete w.r.t.\ $(S,S)$.

\end{example}

\section{Preliminaries}
\label{sec:preliminaries}

\renewcommand{\myparagraph}[1]{%
  \paragraph{#1}}

\myparagraph{\bf Basic notions.}

We assume that the reader is familiar with basics of logic programming, and
 use the standard definitions and notation %
\citep{Apt-Prolog-short,apt/bol}.  A maybe nonstandard notion is
(computed or correct) {\em answer}\/; by this we mean a query to which a
(computed or correct) answer substitution has been applied.
We assume a fixed set of function symbols (including constants) and of predicate
symbols.
\HU stands for the Herbrand universe, and \HB for the Herbrand base;
$ground(P)$ is the set of ground instances of the clauses of a program $P$.
 $\NN$ is the set of natural numbers.
We deal with normal \citep{lloyd87} programs (called also ``general'' 
\citep{apt/bol}), and we usually call them just ``programs''.
A procedure is a set of clauses beginning with the same predicate symbol.

Given a set $S\in\HB$ of ground atoms, 
$\neg S$ will stand for   $\{\neg A \mid A\in S\,\}$.
By a {\em level mapping} we mean a function
$|\ |\colon S\to W$, where $S\subseteq\HB$ and $(W,\prec)$ is a %
 well-ordered set (i.e.\ one in which there does not exist an infinite
 decreasing sequence).

\myparagraph{\bf 4-valued logic.}
This logic plays an auxiliary role in the paper, and
at the first reading the references to it may be skipped.
We employ Belnap's 4-valued logic.  
The truth
values are the subsets of the set of two standard truth values \ttt and \fff.
The subsets $\emptyset,\{\ttt\},\{\fff\},\linebreak[3]\{\ttt,\fff\}$ will be denoted by
$\uuu,\ttt,\fff,\tttfff$, respectively.
The rationale for the four values is that a query, apart from succeeding
(\ttt) and failing (\fff), may diverge (\uuu); additionally a specification
may permit it both to succeed and to fail (\tttfff).
A 4-valued Herbrand interpretation is a subset of $\HB\cup\neg\HB$.
(We will often skip the word ``Herbrand''.)
If such interpretation does not contain both $A$ and $\neg A$
(for any $A\in\HB$) then it is said to be 3-valued.
Consider an interpretation $I\subseteq \HB\cup\neg\HB$, and a ground atom
$A\in\HB$.  The truth value $v\subseteq\{\ttt,\fff\}$ of $A$ in $I$ is
determined as follows:
$\ttt\in v$  iff $A\in I$, and $\fff\in v$  iff $\neg A\in I$.
We may briefly say that $A$ is $v$ in $I$, e.g.\ $p$~is \tttfff and $q$ is
\uuu in $\{p,\neg p\}$.
We write $I,\vartheta\models_4 F$  (respectively $I\models_4 F$)
to state that the truth value of a formula $F$ in an interpretation $I$ and
a variable 
valuation $\vartheta$ (resp.\ all variable valuations) contains~\ttt.
When $I$ is 3-valued, $I\models_3$  stands for $I\models_4$.
The logical operations $\land,\lor$ are defined, respectively, as the glb
(lub) in the lattice $(\{\ttt,\fff\},\preceq_t)$, where
$\fff\preceq_t\uuu\preceq_t\ttt$, %
$\fff\preceq_t\tttfff\preceq_t\ttt$,
and $\tttfff,\uuu$ are incomparable.
The negation is defined by $\neg\ttt=\fff$, $\neg\fff=\ttt$, 
 $\neg\uuu=\uuu$,  $\neg\tttfff=\tttfff$.
  See the work by
  \cite{staerk93.lpind} or \cite{DBLP:journals/jlp/Fitting91} for 
  the treatment of $\leftrightarrow$, and
  further explanations.
{}

\myparagraph{\bf Using standard logic.}
In this work we prefer to avoid dealing with technical details of 4-valued
logic. 
Instead we encode it in the standard 2-valued logic.
Such approach is maybe less elegant, but it seems 
convenient to work within a well-known familiar logic.
We extend the underlying alphabet by adding new predicate symbols, a distinct
new symbol $p'$ for each predicate symbol $p$ (except for $=$).
For the encoding, %
 we use the following notation \citep{drabent/martelli,drabent96}.

Let $\cal F$ be a query, the negation of a query, a clause, or a program.
Then 
$\cal F'$ is  $\cal F$ with $p$ replaced by $p'$ in every
negative literal of $\cal F$  (for any predicate symbol $p$). Similarly,
$\cal F''$ is  $\cal F$ with $p$ replaced by $p'$ in every
positive literal.
If $I\in\HB$ is a (2-valued) Herbrand interpretation then 
$I'$ is the interpretation obtained from $I$ by replacing each
predicate symbol $p$ by $p'$.  
E.g.\ for a clause $C=a\mathop\gets\neg b,c$, we have
$C'=a\mathop\gets\neg b',c$, and  $C''=a'\mathop\gets\neg b,c'$.
Let $I=\{a,c\}$, $J=\{c\}$.  Then
$I\cup J'\models C'$ and
$I\cup J'\notmodels C''$ (as  $J'=\{c'\}$).
Consider a 4-valued interpretation
$I=X\cup\neg(\HB\setminus Y)$ (where $X,Y\in\HB$).
For a query $Q$,
we have
$I\models_4 Q$ iff $X\cup Y'\models { Q}'$ %
and
$I\models_4\neg Q$ iff $X\cup Y'\models\neg Q''$.%

\myparagraph
{\bf Negation in Prolog, Kunen semantics.}
Here we present a brief discussion, for missing details see
e.g.\ the work by \cite{apt/bol} or \cite{Doets.book}.

A natural way of adding negation to Prolog was negation as failure (more
precisely, negation as finite failure, NAFF). 
This means deriving $\neg Q$
if a query $Q$ finitely fails, i.e.\ has a finite SLD-tree without answers.
(In Prolog, $\myprologneg Q$ fails if $Q$ succeeds, and succeeds when $Q$ 
terminates without any answer.)
The corresponding operational semantics for normal programs is
SLDNF-resolution (cf.\ Section \ref{sec:sldnf}).

The appropriate declarative semantics for NAFF was proposed by Kunen
\citep{Kunen87,apt/bol}. (We will call it {\em Kunen semantics},  {\em KS} or
{\em completion semantics}.)
It is defined by means of logical consequences of the program completion $comp(P)$
\citep{clark78} in a 3-valued logic of Kleene%
\footnote{%
The logic may be seen as
Belnap's 4-valued logic without the truth value \tttfff.
\cite{staerk93.lpind} proved that the logical consequences of $comp(P)$
are the same in the 3-valued and the 4-valued logic.
For expressing  Kunen semantics in %
the standard 2-valued logic see e.g.\ the paper by %
\cite{drabent96} or by \cite{staerk93.lpind}
 and references therein.
}%
; %
a query $Q$ (resp.\ its negation $\neg Q$) is a result
 of a program $P$ when $comp(P)\models_3 Q$  (resp.\ $comp(P)\models_3\neg Q$).
So soundness (of some operational semantics) w.r.t.\ KS
means that if %
$Q$ with $P$
succeeds with a computed answer $Q\theta$ then $comp(P)\models_3 Q\theta$, 
and if $Q$ %
fails then $comp(P)\models_3\neg Q$.
Completeness means the implications in the other direction.

Kunen semantics properly describes  NAFF and SLDNF-resolution:
First, SLDNF-resolution is sound w.r.t.\ the Kunen semantics, and
is complete for wide classes of programs and queries.
Also, it can be shown that the only reasons for incompleteness are floundering
(cf.\ Section \ref{sec:sldnf})
or non-fair selection rule \citep{drabent96}.
Moreover, natural ways of augmenting SLDNF-resolution by constructive negation
\citep{stuckey91,drabent95} are sound and complete w.r.t.\ this semantics.

\myparagraph
{\bf The well-founded semantics.}
The suitable declarative semantics for normal programs with negation as
(possibly infinite) failure (NAF) is the well-founded semantics (WFS).
It is given by a specific 3-valued {\em well-founded model} ${\it W F}(P)$ of a
given program $P$ \citep{wfsem,apt/bol}.
WFS is not computable --
it is in general impossible to check whether an infinite tree fails.
An operational semantics for WFS is {SLS-resolution} 
(cf.\ Section \ref{sec:SLS}).
SLS-resolution is sound %
w.r.t.\ WFS, and complete for non-floundering queries
\citep{Ross:SLS,apt/bol}.
NAF and SLS-resolution are approximately implemented by Prolog with tabulation.
The methods proposed in this paper depend on soundness of SLDNF- and
SLS-resolution, but not on their completeness.

\section{Specifications for the context of negation}
\label{sec:specifications}

\noindent
Here we show how approximate specifications from Section \ref{sec:definite}
can be used for normal programs.  
This section is independent from the choice of semantics for negation.
When we do not deal with negation, a specification
describes which atoms may succeed, and which ones have to succeed.
Now we also need to specify which atoms should/may fail.  This would require
four interpretations, which is cumbersome.
However it %
is rather natural to demand that the (ground) atoms
required to succeed are not allowed to fail,
and those not allowed to succeed should fail.
So a pair of interpretations as a specification
will be treated from two points of view, as a
specification for correctness, and for completeness
\citep{DBLP:journals/tplp/DrabentM05shorter}.

\begin{definition}
\label{def.specification}
\hspace*{\parindent}%
An approximate {\bf specification} 
is a pair $(\snf,\st)$ of Herbrand interpretations.

A specification  $(\snf,\st)$  is called {\bf proper} if $\snf\subseteq\st$.
\end{definition}
We usually drop the adjective ``approximate''.
The letters ${\it n f\!,t}$ stand for ``not false'', and ``true'' (cf.\
the 4-valued view of Def.\,\ref{def.4v}).
Comparing with Section \ref{sec:definite},
\snf corresponds to a specification for completeness, and \st to that for correctness.
The diagram illustrates a proper specification $spec=(\snf,\st)$, and the
ground atomic queries that succeed/fail for a program $P$ which is correct and
complete w.r.t.\  $spec$.
\[
%
\input{diagram-approximate-neg}%
%
\]

Considering program correctness,
informal understanding of a specification  $(\snf,\st)$
is that $A\in\st$ means that
$A$ can be (an instance of) an answer of the program; and 
$ A\in\HB\setminus\snf$ means that 
the query $A$ can fail (and $\neg A$ can be an instance of an answer).  
Hence any atom $A\in\st\setminus\snf$ is allowed both to succeed and fail,
and any $A\in\snf\setminus\st$ is neither allowed to succeed, nor to fail.

So, in the notation introduced in Section \ref{sec:preliminaries},
for any answer $Q$ of a program $P$ correct w.r.t.\ $(\snf,\st)$
we have $\st\cup\snf' \models Q'$.  Also, 
if  $Q$ with $P$ fails then  $\st\cup\snf' \models \neg Q''$.
\enlargethispage{2.7ex}

Considering program completeness, specification  $(\snf,\st)$
requires any atom $A\in\snf$ to succeed,
and any $A\in\HB\setminus\st$ to fail.  
So $\st\cup\snf' \models Q''$ means that $Q$ is required 
 to be program answer;
$\st\cup\snf' \models \neg Q'$ means that $Q$ is required to fail.
A non proper specification makes no sense in the context of completeness,
as it requires some atoms to both succeed and fail.

As an example, consider %
specification $(r,r_+)$ for a list
membership predicate from Ex.\,\ref{ex.approx}.
It is proper, and it says that the atoms from $r$ (resp.\,$r_+$) must (may)
succeed (hence those from $\HB\setminus r_+$ cannot succeed).
Taken negation into account, it also states that the atoms from
$\HB\setminus r$ (resp.\  $\HB\setminus r_+$) may (must) fail
(hence those from $r$ cannot fail).

\pagebreak[3]

The following definition provides a 4-valued logic view at approximate specifications.

\begin{definition}
\label{def.4v}
For a specification $spec=(\snf,\st)$, let us define two 4-valued
interpretations:

$I^4(\snf,\st) = \st \cup\neg(\HB\setminus\snf)$
\ \  %
($spec$ treated as a specification for correctness),

$I_4(\snf,\st) =  \snf\cup \neg(\HB\setminus\st)$
\ \ 
($spec$ %
 treated as a specification for completeness).

\end{definition}

\section{Correctness and completeness, Kunen semantics}
In this section we present sufficient conditions for correctness and
semi-completeness for normal programs under Kunen semantics.

\subsection{Program correctness}

\begin{definition}[Correctness, Kunen semantics]
\label{def.correctness}
  A program $P$ is {\bf correct} under KS  %
w.r.t.\ a specification $spec = (\snf\!,\st)$
  if for any query\,\,\,\,\,\makebox[0pt][c]{$Q$\,\,\,}
%
\myenumeratefix
\begin{enumerate}
\renewcommand\labelenumi{(\roman{enumi})}
\renewcommand\theenumi\labelenumi
  \item %
 if $comp(P) \models_3 Q$ then $\st\cup\snf' \models Q'$
\hfil    \quad\ (equivalently, $I^4(\snf,\st) \models_4 Q$),
    \label{def.correctness.i}
  \item %
    \label{def.correctness.ii}
 if $comp(P) \models_3 \neg Q$ then $\st\cup\snf' \models\neg Q''$
\hfil     (equivalently, $I^4(\snf,\st) \models_4\neg Q$).
\end{enumerate}
\end{definition}
For an atomic query $Q = A$, conditions \ref{def.correctness.i}, %
\ref{def.correctness.ii}  reduce respectively to: \
if $comp(P) \models_3 A$ then $\st \models A$, \ and \ 
if $comp(P) \models_3 \neg A$ then $\snf \models\neg A$. 

Soundness of SLDNF-resolution relates this notion of correctness to actual
computations:
if $Q$ is a computed answer for $P$ then $comp(P) \models_3 Q$, and 
if $Q$ finitely fails then  $comp(P) \models_3 \neg Q$.

\begin{definition}
[covered atom, normal programs]
  An atom $A\in\HB$ is {\bf covered} by a clause $C$
  w.r.t.\ a specification $spec = (\snf,\st)$ if there exists a ground
  instance  $A\gets\vec B$ of $C$ such that
  $B\in\snf$ for each positive literal $B$ from $\vec B$, and
  $B\not\in\st$ for each negative literal $\neg B$ from $\vec B$.
  (In other words,  $\st\cup\snf'\models\vec B''$).
  
$A$ is covered w.r.t.\ $spec$ by a program $P$ 
if it is covered w.r.t.\ $spec$ by a clause $C\in P$.

\end{definition}
Speaking informally, $A$ is covered
by clause $C$ if $C$ is able to
produce $A$ out of literals required by $spec$ to be produced by the program.
As an example, we show that $m(a,[c,a])$ is covered by 
$C = m(E,[\myunderscore|T])\mathop\gets m(E,T), \neg m(E,[b])$
w.r.t.\ specification   $(r,r_+)$ from Example \ref{ex.approx}.
The body of the considered ground instance of $C$ is $L_1,L_2$, where
$L_1= m(a,[a])$, $L_2=\neg m(a,[b])$; 
then  $(L_1,L_2)''$ is  
$m'(a,[a]),\neg m(a,[b])$ and $r_+\cup r'\models (L_1,L_2)''$.

The following theorem
\citep{drabent.normal.lopstr2022}
(it also follows from %
the paper by \cite{DBLP:conf/aadebug/Ferrand93})
provides a way of 
proving correctness of normal programs under KS. %

\begin{theorem}[Correctness, Kunen semantics]
\label{th:correctness1}
\myenumeratefix
A program $P$ is correct w.r.t.\ a specification $spec = (\snf,\st)$ 
under KS if
  \begin{enumerate}
  \item \label{correctness1-1}
  $\st\cup\snf'\models P'$, and
  \item \label{correctness1-2}
    each atom $A\in\snf$ is covered by $P$ w.r.t.\ $spec$.
  \end{enumerate}
\end{theorem}
Note that conditions \ref{correctness1-1} and \ref{correctness1-2}
are, in a sense, a natural generalization of
the sufficient conditions for correctness and semi-completeness for
positive programs.
The conditions actually imply correctness of $P$ under \cite{Fitting85} 
semantics
(which is implied by correctness under KS). See the paper by
\cite{drabent.normal.lopstr2022} for a stronger sufficient condition.
Examples of applying the theorem are contained in Examples 
\ref{ex.odd}, \ref{ex.paths.KS} below.

\enlargethispage{2.5ex}

\subsection{SLDNF-resolution}
\label{sec:sldnf}
To generalize semi-completeness for normal programs under %
 Kunen semantics,
we need to refer to an operational semantics.  To avoid including  a
rather lengthy definition of SLDNF-resolution, we make it explicit which aspects
of it will be used here.  In this way the proposed method is sound for any
variant of SLDNF-resolution with the features specified below
(e.g.\ those discussed by \cite{Apt/Doets}).

Firstly, we assume that SLDNF-resolution is sound w.r.t.\ KS 
(cf.\ Section \ref{sec:preliminaries}).
We also assume that, for any program $P$, if $Q$ finitely fails then any its
instance 
$Q\theta$ finitely fails, 
and if $Q$ succeeds with a most general answer (i.e.\ the answer $Q\eta$ is a
variant of $Q$) then any its instance $Q\theta$ succeeds with a most general
answer. 
Some negative literals are distinguished as {\it dealt with} (this may depend
on the program and the selection rule).
If $\neg A$ is dealt with, then $A$ finitely fails or succeeds with a most
general answer.  Moreover,
each ground $\neg A$ for which $A$ finitely fails or succeeds is dealt with.
Also, if $\neg A$ is dealt with, then any its instance $\neg A\theta$ is.

Secondly, we assume that -- given a program $P$, a query $Q$ and a selection
rule $\cal R$ -- the search space for $Q$ contains the %
{\bf main SLDNF-tree} for $Q$ (shortly, main tree)
 as follows. 
(The search space may be a set of trees and successful
derivations with assigned ranks \citep{lloyd87}, 
the SLDNF-tree \citep{Apt/Doets}, etc;
for the approach of %
\cite{lloyd87}, the main tree is the
SLDNF-tree for~$Q$.)

\begin{enumerate}
\item 
The nodes of a main tree $T$ for $P$ and $Q$ via $\cal R$
 are queries.  The root is $Q$, and
 $\cal R$ selects a literal in each non-empty node.  The children of a node
with a positive literal selected are as in an SLD-tree.  Such node is marked
{\em failed} if it has no children.
For a node %
${\vec A},\neg A,{\vec B}$ with a negative literal $\neg A$ selected, one of 
(a),\,(b),\,(c) holds.
\begin{enumerate}
\item
  \label{sldnf.neg.flounder}
  $\neg A$ is not dealt with,
  ${\vec A},\neg A,{\vec B}$ is a leaf of $T$ and is marked {\it floundered},
\item 
  \label{sldnf.neg.fail}
  $\neg A$ is dealt with,
  $A$ succeeds with a most general answer,  ${\vec A},\neg A,{\vec B}$ is
  a leaf of $T$ and is marked  {\em failed},
\item
  \label{sldnf.neg.success}
  $\neg A$ is dealt with,
  $A$ finitely fails, and ${\vec A},\neg A,{\vec B}$ has a single child 
  ${\vec A},{\vec B}$.
\end{enumerate}
A branch of $T$ is called an {\bf SLDNF-derivation} for $Q$.
Such derivation is {successful}, if its last query is empty.

\item
\label{sldnf.ff}
  $T$  (and $Q$) {\bf finitely fails} if $T$ is finite and all its leaves are
  marked {\it failed}.
\item
  $T$ {\bf succeeds} (and $Q$ succeeds) if $T$ contains an empty leaf.  
Let $\theta$ be the
composition of the mgu's (most general unifiers) along a successful branch
$D$ of $T$.
  (Note that there are no mgu's related
  to nodes with a negative literal selected.)  Then $Q\theta$ is 
  the computed {\bf answer}
  of $D$ for $Q$. %

\end{enumerate}

We say that a main tree $T$ flounders if it contains a leaf marked {\em
  floundered}.  
A main tree is  {\bf not diverging} if it is finite, and does not flounder.

Obviously, in practice, it is not enough that the main tree is not diverging,
we need that the whole 
computation terminates, i.e. the whole search space is finite.
For a sufficient condition for termination, see %
the overview by \cite{apt/bol}   %
and the approach of \citet[Section 4.3.5]{DBLP:journals/tplp/DrabentM05shorter}.
 \enlargethispage{.5ex}

\subsection{Completeness of normal programs under the Kunen semantics}
We show that if a program is correct
then it is semi-complete.
The main theorem and the related lemmas follow some ideas of
\cite{DBLP:journals/tplp/DrabentM05shorter}. 
First we generalize the definition of semi-completeness for normal programs
with the Kunen semantics.

\begin{definition}[Completeness, Kunen semantics]
\label{def.complete}
  A program $P$ is {\bf complete for a query} $Q$ under KS
  w.r.t.\ a  specification %
  $(\snf\!,\st)$ if
\myenumeratefix
  \begin{enumerate}
   \item[(i)]  $\st \cup \snf' \models Q''$ implies
     $comp(P) \models_3 Q$,
   \item[(ii)]  $\st \cup \snf' \models \neg Q'$ implies 
     $comp(P) \models_3 \neg Q$.
   \end{enumerate}

\noindent
Program $P$ is {\bf complete} under KS w.r.t. $spec$ if it is complete
under KS w.r.t. $spec$ for any
   query~$Q$.%
\end{definition}

Note that the antecedents of (i), (ii) are respectively equivalent to\,
$I_4(\snf,\st) \models_4 Q$, \ and \  
$I_4(\snf,\st) \models_4\neg Q$.

\begin{definition}
\label{def.semi}
  A program $P$ is {\bf semi-complete} w.r.t.\ a specification $spec$ under KS
if $P$ is complete under KS w.r.t.\ $spec$ for any %
query $Q$ for which there exists a non diverging main
 SLDNF-tree.

A program $P$ is {\bf SLDNF-semi-complete for a query} $Q$
w.r.t.\ a %
 $spec=(\snf,\st)$ if
for any non diverging  main
 SLDNF-tree $T$ for $P$ and $Q$, and any
instance $Q\theta$ of~$Q$
\myenumeratefix
{%
    \begin{enumerate}
    \item[(i)]  $\st \cup \snf' \models Q''\theta$ implies that
     $T$ contains an answer $Q\rho$ more general than $Q\theta$,
    \item[(ii)]  $\st \cup \snf' \models \neg Q'$ implies that
      $T$ finitely fails.
    \end{enumerate}
Program $P$ is {\bf SLDNF-semi-complete} w.r.t.\ $spec$ if $P$
SLDNF-semi-complete for each query $Q$.
\par
}

\end{definition}

In less formal wording, $P$ is SLDNF-semi-complete if each non diverging
main SLDNF-tree produces the results required by the specification.
Note that SLDNF-semi-complete implies semi-complete
(by soundness of SLDNF-resolution).

\begin{example}
[completeness for positive programs versus completeness under KS]
\label{ex.cycles}
  Consider a graph with nodes $a,b,c$ and the set of edges given by a set of
  facts   $G=\{\, e(a,b),  e(b,a) \,\}$.
  Consider a specification $spec=(\st,\st)$ where 
  $\st=G\cup\st_p$ and $\st_p=\{\,p(a,b),  p(b,a), p(a,a),  p(b,b)\,\}$
(so $p$/2 describes which nodes in $G$ are connected).
Consider a program
  $P = G\cup \{\, p(X,Y)\mathop\gets e(X,Y).\
  p(X,Z)\mathop\gets e(X,Y), p(Y,Z). \,\} $.
{\sloppy\par}

 Treated as a positive program, $P$ is correct and complete
  w.r.t.\ $\st$, as $\st$ is its least Herbrand model.
  Under the Kunen semantics, $P$ is
  correct and SLDNF-semi-complete w.r.t.\ $spec$, but not complete.
For instance, $spec$ requires $p(a,c)$ and $p(c,a)$
  to fail. Query $p(c,a)$ finitely fails
(under Prolog selection rule),
 and $comp(P)\models_3\neg p(c,a)$.
  On the other hand,   $p(a,c)$ diverges
 (for any selection rule the SLD-tree is infinite),
  and  $comp(P)\notmodels_3\neg p(a,c)$.%
\enlargethispage{3.2ex}
\footnote
{%
To show that $comp(P)\notmodels_3\neg p(a,c)$,
note that $p(a,c)$ is true in a 3-valued model of $comp(P)$, namely in
$I=\st_1\cup\neg(\HB\setminus\st_1)$ where $\st_1= \st_p\cup \{p(a,c),p(b,c)\}$.
  To show that for any selection rule $p(a,c)$ diverges, note that 
  it does not succeed (as otherwise $P$ is incorrect w.r.t.\ $spec$),
  and it does not finitely fail (as otherwise 
  $comp(P)\models_3\neg p(a,c)$, contradiction).
} %

\end{example}

When considering program completeness, we add a requirement on the underlying
language.
Speaking informally, a sufficient supply of symbols is necessary.
By an {\bf extended language} we mean 
the first order language over an alphabet
containing an infinite set of constants, or a
function symbol not occurring in the considered program and queries.
In a sense, specifications in an extended language do not require too much.
For instance,  when the only function symbol is $a$/0, and a specification
is $spec=(\{p(a)\},\{p(a)\})$, program $P=\{p(a).\}$ is not complete w.r.t.\ 
$spec$.  
This is because $p(X)$ is not an answer of $P$, but
$\{p(a)\}\cup\{p'(a)\}\models p(X)$. This does not hold over an extended
language.

For our main theorem we use two following lemmas.  Their proofs are given
in the supplementary material.

\begin{lemma}\label{lem.terms}
Let $t$ be a term with $k\geq0$ variables.
Let $f$ be a non-constant function symbol not occurring in $t$.  
Alternatively, let $\seq[k]c$ be distinct constants not occurring in $t$.  
Then there exists an instance of  $t$  which is not unifiable
with any term $s$ such that (i) $t$ is not an instance of $s$, and (ii)
$f$ (respectively any of  $\seq[k]c$) does not occur in $s$.

\end{lemma}

\begin{lemma}
\label{lemma.semi-completeness}
Let a normal program $P$ be correct w.r.t.\ a specification $spec=(\snf,\st)$
under KS.
Let  $Q$ be a query.
Assume an extended underlying language.
Let $T$ be a main SLDNF-tree for a query $Q$ and $P$.
If $T$ is non diverging then $T$ produces the results required by $spec$
(i.e.\ $T$ finitely fails if  $\st\cup\snf'\models \neg Q'$, and otherwise
$T$ has a computed answer more general than $Q\theta$ for each
$Q\theta$ such that $\st\cup\snf'\models Q''\theta$).  
\end{lemma}

\begin{theorem}[Semi-completeness, Kunen semantics]
\label{th.semi-completeness}

Let a normal program $P$ be correct w.r.t.\ a specification $spec=(\snf,\st)$
under KS, over an extended language.
Then $P$ is SLDNF-semi-complete (hence semi-complete under KS) w.r.t.\ $spec$.
\end{theorem}

The theorem follows immediately from Lemma \ref{lemma.semi-completeness}.
As shown in supplementary material, 
Lemma \ref{lemma.semi-completeness} holds for ground queries and arbitrary
languages. 
Hence, for any underlying language, correctness (of $P$ w.r.t.\ $spec$) implies 
SLDNF-semi-completeness for ground queries (of $P$ w.r.t.\ $spec$),

\section{Systematic construction of programs, Kunen semantics}
\label{sec:construction.Kunen}
Theorems \ref{th:correctness1} and \ref{th.semi-completeness}
suggest the following method:
To build a program $P$ which is correct and SLDNF-semi-complete (hence
semi-complete) under the Kunen semantics
w.r.t.\ a given specification
$spec=(\snf,\st)$, construct clauses so that
\begin{enumerate}
\renewcommand\labelenumi{(\Alph{enumi})}
\renewcommand\theenumi\labelenumi
\item%
\label{construction.a}
each atom from $\snf$ is covered by some clause $C\in P$
w.r.t.\,$spec$, and
\item%
\label{construction.b}
$\st\cup\snf'\models C'$, for each constructed clause $C\in P$.
\end{enumerate}

\noindent
The program obtained in this way is correct w.r.t.\ $spec$
(by Theorem \ref{th:correctness1}) and SLDNF-semi-complete w.r.t.\ $spec$
(by  Theorem \ref{th.semi-completeness}).  For the program to be useful,
care should be taken to (C) choose the clauses so that the program terminates 
and does not flounder
for queries of interest.  If the constructed program   %
does not diverge for each ground query (under some selection rule) then it is
complete w.r.t.\ $spec$.

\pagebreak[3]
\begin{example}
\label{ex.odd}
\nopagebreak
  We construct a program describing odd natural numbers.  
  First let us provide a specification.
  We will use the
  usual representation, and say that a term $s^i(0)$ ($i\geq0$) is a number.
  So we need that all atoms from 
  $\snfa_o = \{\, o(t)\mid t \mbox{ is an odd number} \,\}$
  are answers of the program.
  We do not bother about terms not representing numbers, so our specification 
  is $spec_o=(\snfa_o,\st_o)$ where
 $\st_o= \{\,o(t)\in\HB \mid \mbox{if $t$ is a number then $t$ is odd} \,\}$.
  Note that $s(t)\in\HB\setminus\st_o$ iff $t$ is an even
  number;  $spec_o$ requires each such $s(t)$  to fail, as expected.

  We need to construct clauses so that each element of   $\snfa_o$ is covered,
  and for each clause $C$ we have $\st_o\cup\snfa_o\!'\models C'$.
  Let us use the fact that a number is odd iff it is a successor of a not odd
  one.
  This leads to $C= o(s(X))\gets\neg o(X)$.
  Note that all elements of $\snfa_o$ are covered by $C$ w.r.t.\  $spec_o$
  (as each odd number is of the form 
  $s(t)$, the corresponding instance of $C$ is $o(s(t))\gets\neg o(t)$,
  $t$ is an even number,  hence $o(t)\not\in\st_o$). 
  Also, $\st_o\cup\snfa_o\!'\models C'$ (as if 
  $\snfa_o\!'\models \neg o'(t)$,  then $t$ is not an odd number, hence
  $s(t)$ is an odd number or not a number, thus $\st_o\models o(s(t))$\,).

So ${\it ODD}=\{C\}$ is the constructed program, correct and semi-complete
w.r.t.\ $spec_o$.
${\it ODD}$ is complete, as it does not diverge for any ground query%
{
(each ground $o(s^i(u))$, where the main symbol of $u$ is not $s$, succeeds or
finitely fails, by induction on $i$).
}
    \end{example}

\begin{example}[Paths in a graph]
\label{ex.paths.KS}
Consider a directed graph $G$ with the edges described by a predicate $e$/2
(possibly given by a set of facts),
 correct and complete w.r.t.
$$
(\st_e,\st_e) \qquad \mbox{where}\qquad
\st_e = \{\, e(t,u)\in\HB \mid
  \mbox{there is an edge  $(t,u)$ in }G\,  \}.
$$
Let us construct a program finding paths between a given pair of nodes.
We know that a naive program for transitive closure may %
miss some expected results for graphs with loops, due to negation by finite
failure,
cf.\ Example \ref{ex.cycles}.
To obtain a usable program, we add an argument with (a list of) nodes that
should not be included in the path.

To construct a specification, let us introduce two auxiliary notions.
By a {\it simple} path we mean one which does not visit a node twice (i.e.\ 
a $[\seq t]$ where $t_i\neq t_j$ for any $i\neq j$).
We say that a path $[\seq t]$ {\it bypasses} a node $e$ when
$e\not\in\{ t_2,\ldots,t_{n-1}\}$.
  It bypasses a list of nodes if it bypasses
every node from the list.  Now consider
\[
\begin{array}{l}
\snfa_p
=\left\{\, p(t,v,s,[\seq u])\in\HB\,\left|
  \begin{array}{@{\:}l@{}}
    s \mbox{ is a simple path in } G \mbox{ from } t \mbox{ to } v    \\
    \mbox{bypassing } [\seq u]
  \end{array}
  \,\right.\right\},
\\
\st_p = \snfa_p\cup \left\{\, p(t,v,s,u)\in\HB \mid 
 u \mbox{ is not a list}
  \,\right\}.
\end{array}
\]
Our current specification for the program is
  $spec_0=(\snfa_p\cup\st_e, \, \st_p\cup\st_e )$.
Informally speaking, it
implies that
answers of the form $p(t,v,s,[\,])$ 
provide (all) the simple paths from $t$ to $v$ in $G$.

Procedure $e$/2 defining the graph is given, 
it remains to construct procedure $p$/4.
For each atom from $\snfa_p$ we need a clause covering it.
Consider an atom $A_1 = p(t,t,s,u)\in\snfa_p$.
Note that $s=[t]$, as the path is simple.
$A_1$ is covered by a fact $C_1= p(X,X,[X],\myunderscore)$
(which satisfies \ref{construction.b}, as $C_1'=C_1$ and $\st_p\models C_1$).

Consider an atom $A_2 = p(t,v,s,u)\in\snfa_p$, where $t\neq v$.  
Then $s=[t|s_1]$ for a simple path $s_1$ from a node $t_1$ ($t_1\neq t$) to $v$,
so that  $(t,t_1)$ is an edge in $G$.
As $s$ bypasses $u$,   $t_1\not\in u$ and $s_1$ bypasses $u$.
Also, $s_1$ bypasses $t$, as otherwise $s$ is not simple.

A candidate for a ground clause covering $A_2$ could be
$C_{20}=p(t,v,[t|s_1],u)\gets e(t,t_1), \linebreak[3] p(t_1,v,s_1,[t|u]).$
However, it is easy to see that it would violate \ref{construction.b}
(and lead to an incorrect program).
We need to assure that $t\neq t_1$ and  $t_1\not\in u$, in other words
$t_1\not\in [t|u]$.
For this it is convenient to employ a list membership program $M$ from Example
\ref{ex.member},
with specification $spec_m=(r,r_+)$, %
 where
  $r= \{\, m(e_i,[\seq e])\in\HB \mid 1\leq i\leq n \,\}$,
 and
  $r_+= r\cup \{\, m(e,u)\in\HB \mid u \mbox{ is not a list} \,\}$.
Our specification becomes  %
\[
spec=(\snf,\st), \quad\mbox{where}\quad
 \snf = \snfa_p\cup\st_e\cup r, \quad
 \st = \st_p\cup\st_e\cup r_+.
\]
Consider a clause
\[
C_2 = p(T,T1,[T|S],U)\gets e(T,T2),\neg {\it m(T2,[T|U])},
p(T2,T1,S,[T|U]).
\]
and its ground instance
$$
C_{21}= \underbrace{p(t,v,[t|s_1],u)}_H   %
\gets \underbrace{e(t,t_1)}_{B_1}, \neg \underbrace{\it m(t_1,[t|u])}_{B_2},
\underbrace{p(t_1,v,s_1,[t|u])}_{B_3}.
$$
Condition \ref{construction.a} holds:
Assume  $H=A_2$ and $t_1$ as above.  Then
$C_2$ covers $A_2$ w.r.t.\ $spec$, as
$B_1,B_3\in\snf$ and $B_2\not\in\st$, due to the facts
described above.
(E.g.\ $B_3\in\snf$ as $s_1$ is a simple path from $t_1$ to $v$; and $s_1$
bypasses $t$ and the nodes in $u$.)

To show that $C_{21}$ satisfies condition \ref{construction.b}
(i.e.\,$\st\cup\snf'\models C_{21}'$ for each ground instance $C_{21}$ of $C_2$),
assume that $B_1,B_3\in\st$ and $B_2\not\in\snf$. 
We have to show that  
$H\in\st_p$.  This is immediate if %
 $u$ is not a list.
Otherwise  $[t|u]$ is a list.
Hence from $B_3\in\st$ we obtain that 
$s_1$ is a simple path from $t_1$ to $v$ bypassing $[t|u]$.
So $[t|s_1]$ is a path from $t$ to $v$ (by $B_1\in\st$).
From $B_2\not\in\snf$ we obtain (as $[t|u]$ is a list) that $t_1\not\in [t|u]$.
So $t_1\neq t$ and thus path $[t|s_1]$ is simple.
Also $[t|s_1]$ bypasses $u$, as the first element $t_1$ of $s_1$ does not occur
in $u$ and $s_1$ bypasses $u$. 
Hence $H\in\st_p$.  %

Thus the constructed program  ${\it PATH} = \{C_1,C_2\}\cup M$ is
correct and SLDNF-semi-complete
w.r.t.\ $spec$,

We explain that, under the Prolog selection rule,
the main SLDNF-tree for {\it PATH} and a query $Q_0=p(t_0,t_1,s,u_0)$ (where $t_0,u_0$
are ground) does not diverge.  Whenever $\neg m(\ldots)$ is
selected, its arguments are ground, so the subsidiary tree (for
$ m(\ldots)$) is finite. 
This implies lack of floundering.
On each path in the main tree, every third element is of the form
$Q_{3i}=p(t_i,t_1,s_i,u_i)$ ($i\geq0$), where $t_i\not\in u_i$ for $i>0$,
and  $u_i=[t_{i-1},\ldots,t_0|u_0]$.
So $t_{i-1},\ldots,t_0$ are distinct nodes of $G$, which is finite.
Thus every such path is finite, the tree does not diverge, and thus
  {\it PATH} is complete w.r.t.\ $spec$.

\end{example}
\section{The well-founded semantics}

This section presents sufficient conditions for program correctness and
semi-completeness under the well-founded semantics.

\subsection{Program correctness, the well-founded semantics}

A definition of correctness under WFS can be obtained
from that of Def.\,\ref{def.correctness} by replacing
 $comp(P) \models_3$\,  by  $\it W F(P) \models_3$.  Here is an equivalent and
simpler formulation.

\begin{definition}\rm
  A program $P$ is {\bf correct} under the well-founded semantics w.r.t.\ a
  specification $spec = (\snf,\st)$ when
  ${\it W F}(P)\subseteq I^4(spec)$.
\quad
In other words, ${\it W F}(P)\subseteq \st\cup \neg(\HB\setminus\snf)$.

  A program $P$ is {\bf complete} under the well-founded semantics w.r.t.\ a
  specification $spec = (\snf,\st)$ when
  $ I_4(spec)\subseteq{\it W F}(P)$.
\quad
In other words
$\snf \cup\neg(\HB\setminus\st)\subseteq{\it W F}(P)$.
\end{definition}
\begin{theorem}[Correctness, WFS]
\label{th:correctness.wfs}
\rm
A program $P$ is correct w.r.t.\ a specification $spec=(\snf,\st)$
under the well-founded semantics provided that
\myenumeratefix
    \begin{enumerate}
  \item
    \label{correctness31}
    $\st\cup\snf'\models P'$, and
  \item
    \label{correctness32}
    there exists a level mapping $|\ |\colon \snf\to W$ such that 
    \\
    for each $A\in\snf$ there exists  $A{\gets}\vec L \in ground(P)$
    such that 
\vspace{-.5\medskipamount}
    \begin{enumerate}
      \item
    \label{correctness321}
    $A$ is covered by %
    $A{\gets}\vec L$ (i.e.\ 
       $\st\cup\snf'\models\vec L''$),  %
      \item
    \label{correctness322}
        for each positive literal $L$ from $\vec L$, $|L|\prec |A|$.
    \end{enumerate}
    \end{enumerate}
\end{theorem}
The theorem is due to \citeN{ferrand/deransart92}
(they use different terminology, among others program
correctness is divided into ``partial correctness'' and ``weak completeness'').
An informal explanation for condition \ref{correctness322}
is that for each $A\in\snf$ it implies existence of a kind of a proof tree of
finite height with root $A$. %

Answer set programming (ASP) is outside the scope of this paper.
We however mention the following property
(see the supplementary material for a proof and example).

\begin{proposition}
\label{propo.stable}
If condition \ref{correctness31} %
of Theorems \ref{th:correctness1}, \ref{th:correctness.wfs} %
holds for a program
$P$ then for each stable model $I$ of $P$ 
if $\snf\subseteq I$ then $I\subseteq \st$.
 \vspace{-1ex}   
\end{proposition}

\subsection{SLS-resolution}
\label{sec:SLS}

SLS-resolution \citep{Prz:JAR89,Ross:SLS,apt/bol}
is an operational semantics for WFS.
Instead of its definition, we present its aspects of interest.

We assume that SLS-resolution is sound.  This means that, for a program $P$,
if a query $Q$ fails then $\it W F(P)\models_3\neg Q$, and if $Q$ succeeds with
answer $Q\theta$ then  $\it W F(P)\models_3 Q\theta$.
As previously, we assume that for any program $P$,
if $Q$ fails or succeeds with a most general answer,
then the same holds for any its instance $Q\theta$.

We also assume that -- given a program $P$, a query $Q$, and a selection
rule $\cal R$ -- the search space for $Q$ contains the %
{\bf main SLS-tree} (shortly, main tree), which is similar to a main
SLDNF-tree of Section \ref{sec:sldnf}.  The only difference is that a failed
tree is not required to be finite.  
So the requirements are as previously, with 
``is finite'' dropped from condition \ref{sldnf.ff},
``finitely fails'' replaced by ``fails'', and ``SLDNF-'' by ``SLS-''.
Also, $T$ is
not diverging if $T$ does not flounder.

 \enlargethispage{1.9ex}  
For example,  the main SLS-tree
for program $P_1=\{\,p(s(X))\mathop\gets p(X).\,\}$ and query $p(X)$ fails;
it consists of a single infinite branch.  
For
 $P_2=\{\,a\mathop\gets\neg a.\,\}$ and query $a$, the main SLS-tree $T$
consists 
of two nodes $a,\neg a$, and the leaf $\neg a$ is marked {\it floundered}.
(The ``global SLS-tree''\! 
\citep[Def.\,9.9--9.11]{Ross:SLS,apt/bol}
has $T$ as its root.)

\subsection{Program completeness under the well-founded semantics}

We begin, as previously, with introducing some technical notions.
\pagebreak[3]
\begin{definition}
\label{def.complete.wfs.for}
\nopagebreak
  A program $P$ is {\bf complete for a query} $Q$ under WFS 
  w.r.t.\ a  specification $(\snf,\st)$ if
\myenumeratefix
  \begin{enumerate}
  \item[(i)]  $\st \cup \snf' \models Q''$ implies
    $\it W F(P) \models_3 Q$,
  \item[(ii)]  $\st \cup \snf' \models \neg Q'$ implies 
    $\it W F(P) \models_3 \neg Q$.
  \end{enumerate}
\end{definition}
Note that a program is complete for any ground query iff it is complete 
(under WFS, w.r.t.\ to a given specification).

\begin{definition}
\label{def.semi.wfs}
  A program $P$ is {\bf semi-complete} w.r.t.\ a specification $spec$ under WFS
if $P$ is complete under WFS w.r.t.\ $spec$ for any %
query $Q$ for which there exists a non diverging main SLS-tree.
A program $P$ is {\bf SLS-semi-complete for a query} $Q$
w.r.t.\ a %
 $spec=(\snf,\st)$ if
for any non diverging main SLS-tree $T$ for $P$ and $Q$, and any
instance $Q\theta$ of $Q$
\myenumeratefix
{%
    \begin{enumerate}
    \item[(i)]  $\st \cup \snf' \models Q''\theta$ implies that
     $T$ contains an answer $Q\rho$ more general than $Q\theta$,
    \item[(ii)]  $\st \cup \snf' \models \neg Q'$ implies that
      $T$ fails.
    \end{enumerate}
Program $P$ is {\bf SLS-semi-complete} w.r.t.\ $spec$ if $P$ is
SLS-semi-complete w.r.t.\ $spec$ for each query $Q$.
\par
}
\end{definition}

By soundness of SLS-resolution, SLS-semi-completeness implies semi-completeness
(of $P$ w.r.t.\ $spec$ under WFS).

As explained in  the supplementary material, 
Lemma \ref{lemma.semi-completeness} holds also for WFS (with KS
replaced by WFS, and SLDNF by SLS).  From the lemma 
it immediately follows:

\begin{theorem}
  [Semi-completeness, the well-founded semantics]
\label{th.semi-completeness.wfs}
Let a normal program $P$ be correct w.r.t.\ a specification $spec=(\snf,\st)$
under WFS, over an extended language.
Then $P$ is SLS-semi-complete (hence semi-complete under WFS) w.r.t.\ $spec$.
\end{theorem}

As in the case of Theorem \ref{th.semi-completeness},
if we drop the requirement on the language,
then SLS-semi-completeness for ground queries is implied.

\section{Construction of programs, the well-founded semantics
}
\label{sec:construction.WFS}

Theorems \ref{th:correctness.wfs}, \ref{th.semi-completeness.wfs}
suggest the following method:
To build a program $P$ correct and SLS-semi-complete (hence semi-complete)
under WFS w.r.t. a
specification $spec=(\snf,\st)$, choose a level mapping  $|\ |\colon \snf\to W$
to a well-ordered set $(W,\prec)$, and
construct clauses so that 
\enlargethispage{.5ex}\vspace{-.5ex}
\begin{enumerate}
\renewcommand\labelenumi{(\Alph{enumi})}
\renewcommand\theenumi\labelenumi
\item%
each atom $A\in\snf$ 
  \begin{enumerate}
    \renewcommand\labelenumii{(\arabic{enumii})}
    \renewcommand\theenumii\labelenumii
  \item 
    \label{construction.wfs.a}
    is covered by a ground instance
    $A\gets\vec L$  of a constructed clause~$C$, so that
  \item
    \label{construction.wfs.b}
    for each positive literal $L$ from $\vec L$, $|L|\prec|A|$, %
  \end{enumerate}

\item 
\label{construction.wfs.c}
    $\st\cup\snf'\models C'$, for each constructed clause $C$.
\end{enumerate}

\noindent
For the obtained program to be useful, care should be taken to (C)~choose the
clauses so that the program does not diverge for the queries of interest.
If the constructed program does not diverge for each ground query
(under some selection rule) then it is complete w.r.t.\ $spec$.

\begin{example}
We
construct a procedure $p$/2 finding, under the well-founded semantics,
whether two nodes of a given finite directed graph $G$ are connected.  
The specification is %
\[
spec=(\snf,\st), \   \mbox{where} %
\begin{array}[t]{l}
\st = \snf = \st_e\cup\st_p,    %
\\
  \st_e  =  \{\, e(t,u)\in\HB \mid
  \mbox{there is an edge  $(t,u)$ in }G\,  \},
\\
  \st_p = %
  \left\{  p(t,u)\in\HB \mid
    \mbox{there is a path from  $t$ to $u$ in $G$\,} %
  \right\}.
\end{array}
\]
Treating $\st_e$ as a set of facts provides a procedure $e$/2 correct and
complete w.r.t. $spec$.  To construct $p$/2 let us first choose the level
$|p(t,u)|$ to be the length of the shortest path from $t$ to $u$
(for any $p(t,u)\in\st_p$), and let
$|e(t,u)|=0$ (for any $e(t,u)\in\st_e$).
Now we need to construct clauses so that each atom  $p(t,u)\in\st_p$ is 
covered.
A border case is $t=u$, for this we use a fact $C_1=p(X,X)$
(for which
\ref{construction.wfs.a}, \ref{construction.wfs.b}, \ref{construction.wfs.c}
obviously hold).
For $t\neq u$, there exists an edge $(t,s)$ to a node $s$ connected with $u$.
This suggests a clause 
\vspace{-1ex}
\[  
C_2 \ = \ p(X,Z) \gets e(X,Y), p(Y,Z).
\]
Note that \ref{construction.wfs.c} holds for $C_2$.
Consider a path of length $|p(t,u)|$ from $t$ to $u$,
and its first edge $(t,s)$.  Obviously there exists in $G$ a path
from $s$ to $u$ of length $|p(t,u)|-1$ ; so $p(s,u)\in\st_p\subseteq\snf$.
Also, $e(t,s)\in\st_e\subseteq\snf$.  Thus
$p(t,u)$ is covered by an instance $C_{20}=p(t,u)\gets e(t,s),p(s,u)$ of $C_2$;
 \ref{construction.wfs.b} holds for $C_{20}$ as $|p(s,u)|<|p(t,u)|$.

The constructed program ${\it PATH2}=\st_e\cup\{C_1,C_2\}$
 is correct and SLS-semi-complete w.r.t.\ $spec$.
It is also complete w.r.t.\ $spec$, as diverging main SLS-trees
for atomic queries do not exists,
due to lack of negation in the program.
In particular, $p(t,u)$ would fail for any not connected nodes $t,u$, even if
the graph contains loops.
Note that for such graphs {\it PATH2} is not complete w.r.t.\ $spec$ under
the Kunen semantics.

\end{example}

\section{Conclusions}

This paper introduces two methods of constructing normal programs which are
correct and semi-complete w.r.t.\ given specifications.  
The methods deal, respectively, with 
the 3-valued completion semantics of Kunen, and the
well-founded semantics.  (The former corresponds to Prolog, the latter to
Prolog with tabulation.)
The approach is declarative, one reasons in terms of
programs seen as logical formulae, and abstracts from operational semantics.
The methods are simple,
 each is described in some 1/4 page
(cf.\ the beginnings of Sections \ref{sec:construction.Kunen}, 
\ref{sec:construction.WFS}). 
They however do not formalize how to assure non-divergence of the constructed
program.
The methods are directly derived from sufficient conditions for correctness
and semi-completeness.
The approach uses, in a sense, a 4-valued logic, however everything is
encoded in the standard logic.  
In particular, a specification is a pair of 4-valued Herbrand
interpretations represented as two 2-valued ones.  An important feature is
that the specifications are approximate, they do not exactly describe the
program semantics.

This work does not deal with ASP (answer set programming); roughly speaking
because our specifications 
describe single literals, while in ASP the outcome of programs
are sets of literals (stable models).
However, sufficient conditions used here lead to a maybe useful
characterization
of stable models (Proposition \ref{propo.stable}).

The simplicity of the methods is partly due to introducing semi-completeness
(i.e.\ completeness for non diverging queries).  This divides completeness
proofs into those for semi-completeness and non-divergence.
Then we get semi-completeness for free, as it is implied by correctness.
This (possibly surprising) property holds for both semantics; note
however the dual understanding of specifications, see Section
\ref{sec:specifications}. 
Another interesting fact is that the method for Kunen semantics is basically
the same as that for the case without negation \citep{drabent.tplp18}.

In the author's opinion, the proposed methods can be used, possibly at some
informal level, in practical every-day programming.  They should be useful 
in teaching logic programming.
The employed sufficient conditions for correctness and semi-completeness
are of separate interest.  They show how to reason declaratively about
program properties.  Again, they can be used in practical programming, and in
teaching logic programming.
For logic programming to be considered a useful declarative programming
paradigm, it is necessary to have declarative methods for reasoning about
programs, and for constructing programs.  This work contributes to providing
such methods. 

\bibliographystyle{acmtrans}
\bibliography{bibshorter,ref,bibmagic,bibpearl}

\newpage
\phantomsection
\addcontentsline{toc}{section}{Supplementary material}

\begin{center}
\vspace*{-3ex}  
  {\LARGE\bf\it  Supplementary material
  \\[.5ex]}
\normalsize
  for  \\[.5ex]
  \large
  {\it On systematic construction of correct logic programs} \\[1ex]
  \normalsize
  2025-08-20 
\end{center}

\appendix  

\bigskip

\noindent
This appendix contains the proofs that could not be included in the paper
due to lack of space.  Then it presents 
Example \ref{ex.propo.stable} of applying Proposition \ref{propo.stable},
and a brief discussion on constructing programs for non-proper
specifications, illustrated by Example \ref{ex.nonproper}.

\bigskip
\noindent{\it Lemma \ref{lem.terms}}\\
Let $t$ be a term with $k\geq0$ variables.
Let $f$ be a non-constant function symbol not occurring in $t$.  
Alternatively, let $\seq[k]c$ be distinct constants not occurring in $t$.  
Then there exists an instance of  $t$  which is not unifiable
with any term $s$ such that (i) $t$ is not an instance of $s$, and (ii)
$f$ (respectively any of  $\seq[k]c$) does not occur in $s$.

\medskip

\begin{proof}
      Let $\seq[k]v\ (k \ge 0)$ be the variables of $t$.
Let $\seq[k]u$ be distinct ground terms with the main symbol not occurring
in $t,\seq t$.
(They are  $\seq[k]c$, or $k$ distinct terms with the main symbol $f$.)
  Consider a substitution $\theta=\{v_1/u_1,\ldots,v_k/u_k\}$.
  We show that $t\theta$ is not unifiable with any $s$ 
  satisfying the conditions of the theorem.
  Consider such $s$ and
  assume that there exists an mgu $\sigma$ of $t\theta$ and $s$. %
  As $t\theta$ is ground,  $t\theta = s\sigma$.
  Terms $\seq[k]u$ occur (as subterms) in $\sigma$ (since they occur in
  $t_i\sigma$, but their main symbols do not occur in~$s$).
  Let us replace each maximal occurrence of term $u_j$  in $\sigma$  by the
  variable $v_j$, obtaining $\sigma_1$.
  (By a maximal occurrence of $u_j$ we mean one which does not occur in an occurrence
  of $u_l$ ($l\neq j$) in $\sigma$.)
  Remove all the pairs of the
  form $v_j/v_j$ from $\sigma_1$, obtaining a substitution $\sigma_2$.
  Then $t_i\sigma_2$ is $t\theta = s\sigma$ with each maximal occurrence of
  any $u_j$ replaced by $v_j$. 
  Hence  $t = t_i\sigma_2$, contradiction with the
  assumption of the lemma.  So $t\theta$ is not unifiable with $s$.
\end{proof}

For the proof %
of Lemma {\ref{lemma.semi-completeness}} 
we generalize to SLDNF- (and SLS-) derivations 
the notion of a {\em lift} of an SLD-derivation.
We add to the original definition by \citet[Def.\,3.14]{Apt-Prolog-short} that
a lift of a derivation step $(\vec X,\neg A,\vec Y)\eta$ $\Rightarrow$ 
$(\vec X,\vec Y)\eta$ \ is \
$(\vec X,\neg A,\vec Y)$ $\Rightarrow$ $(\vec X,\vec Y)$,
provided that $\neg A$ is dealt with and $A$ finitely fails 
(or fails, in the case of SLS-derivation).
Otherwise, query $(\vec X,\neg A,\vec Y)$ is the last query of the lift, and
we say that the lift is {\em floundered}.
Note that each query $R_i$ of an SLDNF- (SLS-) derivation $D=R_1,R_2,\ldots$ is an
instance of the query $S_i$ of its lift $D_l=S_1,S_2,\ldots$, provided that
$D_l$ is not floundered.
Also, a lift $D_l$ of a derivation $D$ may be longer than $D$.
(A clause head may be unifiable with the selected atom $A$ of $S_i$, but not
with the selected atom $A\theta$ of $R_i$.)
On the other hand, a floundered lift  $D_l$ of $D$ may be shorter than $D$.

We have a following lemma,
with a proof basically the same as that 
of Lifting Corollary 2.32 of \cite{Apt-Prolog-short}.
\begin{lemma}[Lifting]
\label{lemma.lifting}
If $D,D_l$ are SLDNF- (or SLS-) derivations respectively for $Q\rho,\, Q$, such that
  $D_l$ is a non floundered lift of $D$, and  $D$ is successful then
  $D_l$ is successful and the answer of $D$ is an instance of the answer of
  $D_l$.   
\end{lemma}

\noindent{\it Lemma \ref{lemma.semi-completeness}}\\
Let a normal program $P$ be correct w.r.t.\ a specification $spec=(\snf,\st)$
under KS.
Let  $Q$ be a query.
Assume an extended underlying language.
Let $T$ be a main SLDNF-tree for a query $Q$ and $P$.
If $T$ is non diverging then $T$ produces the results required by $spec$
(i.e.\ $T$ finitely fails if  $\st\cup\snf'\models \neg Q'$, and otherwise
$T$ has a computed answer more general than $Q\theta$ for each
$Q\theta$ such that $\st\cup\snf'\models Q''\theta$).  

\medskip
\begin{proof}%
Consider a query $Q$ for which there exists a non diverging main SLDNF-tree
$T$. 

1. Assume that $\st\cup\snf'\models \neg Q'$. Suppose that there is a success
leaf in $T$.  By correctness of $P$ w.r.t.\ $spec$,
$\st\cup\snf'\models Q'\theta$
(for the computed answer $Q\theta$
corresponding to the leaf).  Hence $\st\cup\snf'\notmodels \neg Q'$,
contradiction.  Thus $Q$ finitely fails.

2. Assume that  $\st\cup\snf'\models Q''\theta$.
Let $\seq Q$ ($n\geq0$) be the answers of $T$.
We need to show that  $Q\theta$ is an instance of some of them.
Assume the contrary.
By Lemma \ref{lem.terms}
there exists an instance  $Q\rho=Q\theta\varphi$ which is not unifiable
with any of $\seq Q$.

Consider a main SLDNF-tree $T_\rho$ for $Q\rho$ under the same selection rule
as $T$ (more precisely, the rule is lifting invariant,
cf.\ the book by \citet[Def.\,5.43]{Doets.book}).
Then for each branch $D_\rho$ of $T_\rho$ there exists a branch
$D$ of $T$ that is a non floundered lift of $D_\rho$.
This is because for each $\neg A$ selected in $T$, and each instance
$A\rho_1$ of $A$, both
$A$ and $A\rho_1$ finitely fail, or both succeed with a most general answer.

Thus, by Lemma \ref{lemma.lifting},
any answer of $T_\rho$ is an instance of some answer $Q_i$ of $T$.
Thus such answer does not exist,
as it is an instance of $Q\rho$, and $Q\rho$, $Q_i$ are not unifiable.
Also,  $T_\rho$ is finite, as $T$ is.
Hence $T_\rho$ finitely fails, and
$\st\cup\snf'\models\neg Q''\!\rho$, by correctness of $P$.   Contradiction
(with $\st\cup\snf'\models Q''\theta$, as $Q''\!\rho$ is an instance of
$Q''\theta$).
\end{proof}

Note that Lemma \ref{lemma.semi-completeness}
holds for an arbitrary language, if we consider only ground  $Q\theta$.
This is because in the proof 
the restriction on the language is employed to apply 
Lemma \ref{lem.terms}, and for a ground  $Q\theta$ that lemma 
is not needed
(as in this case $Q\rho=Q\theta$, and $Q\rho$ is not an instance of $Q_i$
iff if it is not unifiable with  $Q_i$).

Note also that Lemma  \ref{lemma.semi-completeness}, with KS replaced by WFS,
and SLDNF by SLS, holds for the well-founded semantics and main SLS-trees.   
The proof is the same, with obvious modifications -- in particular allowing
the main tree $T$ to be infinite.

\bigskip

\noindent{\it Proposition \ref{propo.stable}}\\
Let     $\st\cup\snf'\models P'$ for
a program $P$ and a specification $spec=(\snf,\st)$.
Then for each stable model $I$ of $P$,
if $\snf\subseteq I$ then $I\subseteq \st$.

  \medskip\noindent
  {\sc Proof}.
We begin with some definitions \citep{apt/bol}.
  Let $p o s(Q)$
 (resp.\ $neg(Q)$) be the sequence of the
  positive (negative) literals of a query $Q$.
  By $\M_{\it PP}$ we denote the least Herbrand model of a positive program 
$\it PP$.
The Gelfond-Lifschitz reduct of $P$ relative to a (two-valued) interpretation
 $I\subseteq\HB$ is
  $$
 {H(P,I)} = \{\,H\mathop\gets p o s(Q)\mid\linebreak[3] H\mathop\gets Q\in ground(P),\ 
  I\models neg(Q) \,\}
  .$$
  $I$ is a stable model of $P$ when  $I=\M_{H(P,I)}$.
  Note that $I'\subseteq I$  implies  $H(P,I)\subseteq H(P,I')$, hence
  $\M_{H(P,I)}\subseteq\M_{H(P,I')}$.

  Assume  $\snf\subseteq I$.  To show that $I\subseteq\st$ we show that
  (1) $I\subseteq \M_{H(P,\snf)}$ and (2) $\M_{H(P,\snf)}\subseteq \st$.
  Claim (1) follows immediately from $I=\M_{H(P,I)}$ and $\snf\subseteq I$.

  Consider a clause $C$ from $H(P,\snf)$.
  It is obtained from a $H\mathop\gets Q\in ground(P)$, where 
  $\snf\models neg(Q)$  and $C=H\gets p o s(Q)$.
  As $\st\cup\snf'\models H\gets Q'$, we have $\st\models C$.
  We showed that $\st\models H(P,\snf)$.
  Thus 
  the least Herbrand model is a subset of $\st$, i.e.\
 $\M_{H(P,\snf)}\subseteq \st$. $\Box$

\begin{example}[Applying Proposition  \ref{propo.stable}]
\label{ex.propo.stable}
  Consider a program $P=\{\,a\mathop\gets\neg b.\ b\mathop\gets\neg a.\,\}$
  and $\snf=\{a\}$.
  $\st\cup\snf'\models P'$ holds for $\st=\snf$.  
By Proposition \ref{propo.stable} a stable model of $P$ containing $a$ does not
contain anything more.

Consider a program $P_1=\{\,a\mathop\gets\neg b.\ b\mathop\gets\neg c.\ 
                c\mathop\gets\neg a.\,\}$.
A stable model of $P_1$ must be nonempty, as $\emptyset$ is not a model of
$P_1$).  To consider stable models containing $a$ let $\snf=\{a\}$.
$\st_1\cup\snf'\models P_1'$ holds for $\st_1=\{a,b\}$.  
So $I\subseteq\{a,b\}$ for any stable model $I$ of $P_1$ containing $a$.
Consider $\snf_2=\st_1$. Then $\st_2\cup\snf'_2\models P_1'$ holds for 
$\st_2=\{b\}$. Thus no stable model $I\supseteq\{a,b\}$ exists
(as $\{a,b\}\subseteq I\subseteq\{b\}$, contradiction).
  Hence if a stable
model $I$ contains $a$ then $I=\{a\}$.  However this is not a model of $P_1$.
Thus there exists no stable model of $P_1$ containing $a$.

By symmetry, the same holds for $b$ and $c$.  Hence $P_1$ does not have a
stable model.
\end{example}

\paragraph{Non-proper specifications.}
Note that the program construction approaches from the main paper do not
require specifications to be proper.  This may seem wrong, as
completeness w.r.t. a non-proper specification does not make sense.
However semi-completeness does. 
Completeness w.r.t.\ $spec$ requires atoms from  $\snf\setminus\st$ to
both succeed and fail, however correctness %
w.r.t.\  $spec$ requires that
these atoms diverge.  Thus semi-completeness
w.r.t. $spec$ does not deal with these atoms (provided the program is correct
w.r.t.\ $spec$).  For non-proper specifications the proposed methods provide,
as they should, programs that are correct and semi-complete w.r.t.\ the given specifications.

\begin{example}[Program construction for a non-proper specification]
\label{ex.nonproper}
Consider a two person game, with a graph of positions given by a relation
$mo v$/2. 
We may assume that it is given by a set of facts $S_{mo v}$.
A player loses, if she cannot move; a player wins if her opponent loses.
So the sets $W$ of the winning positions and $L$ of the losing ones satisfy
\begin{equation}
\label{eq.WL}
\begin{array}{l}
  L= \{\, u\in\HU \mid \forall_v\
       mo v(u,v)  \Rightarrow v\in W\,\}, \\
  W= \{\, u\in\HU \mid \exists_v\
       mo v(u,v)  \land v\in L\,\}, \\
  L\cap W = \emptyset       
\end{array}  
\end{equation}
(For simplicity, $L$  also includes those
 ground terms that are not nodes
in the graph of the game.)

Let us construct a program describing winning and losing positions.
We may employ a single predicate $w$/1; $w(u)$ should succeed for $w\in W$,
fail for $u\in L$, and diverge otherwise.  This leads to a specification
\begin{equation}
\label{spec.WL}
\begin{array}{ll}
spec=(\snf,\st), \mbox{ where }
&
  \st = S_{mo v} \cup \{\, w(u) \mid u\in W \,\}, \\
&  \snf= S_{mo v}\cup \{\, w(u) \mid u\not\in L \,\}.
\end{array}  
\end{equation}
\enlargethispage{.4ex}
So a ground atom $w(u)$ is allowed by $spec$ to succeed if $u\in W$,
and allowed to fail if $u\in L$.
When $L\cup W\neq\HU$ then $spec$ is not proper
(e.g.\ when
$\HU=\{a,b\}$, $S_{mo v}=\{mo v(a,b),mo v(b,a)\}$, $W=L=\emptyset$).
Note that property (\ref{eq.WL}) does not uniquely define $W$ and $L$, so
$spec$ is not uniquely defined either.

We construct a (single) program correct and semi-complete under WFS
w.r.t.\ any  $spec$ of (\ref{spec.WL}).%
\footnote{
  The program correctness implies the following.  %
  If $w(u)$ is an answer  (respectively, fails), %
  then $w(u)\in\st$ ($w(u)\not\in\snf$), for {\em every} such specification.
  If $w(u)\in\snf\setminus\st$ for {\em some} such specification, then
  $w(u)$ diverges.
}

Consider a $w(u)\in\snf$.  Then $u\not\in L$.  So there is a move
$mo v(u,v)$ such that $v\not\in W$.
This suggests a clause covering $w(u)$, namely
\[
C= w(U)\gets mo v(U,V), \neg w(V).
\]
Indeed, its ground instance %
 $w(u)\gets mo v(u,v),\neg w(v)$ 
covers $w(u)$, as $mo v(u,v)\in\snf$ and $w(v)\not\in\st$.

For the required level mapping we may choose $|w(t_1)|=1$,
$|mo v(t_1,t_2)|=0$ (for any $t_1,t_2\in\HU$).
It remains to show that $\st\cup\snf'\models C'$.  
Take a ground instance $C'_0=w(u)\gets mo v(u,v),\neg w'(v)$ of $C'$,
and assume $mo v(u,v)\in\st$ and $w'(v)\not\in\snf'$.
Hence $v\in L$, thus $u\in W$.  So $w(u)\in\st$, and thus
$\st\cup\snf'\models C'_0$.   

The constructed program ${\it WIN}=\{C\}$ is correct and semi-complete
w.r.t.\ $spec$.  Note that this holds under both semantics: WFS and KS.

\smallskip\noindent
We are interested in  ${\it WIN}$  being complete for any query $w(u)$ where
$u$ is a winning or a losing position.  So we need to show that  ${\it WIN}$
does not diverge for such queries.
Hence $L, W$ should be exactly the sets of losing (resp. winning) positions. 
For this let us provide
possibly transfinite sequences  $L_\alpha$,   $W_\alpha$,
with $L=\bigcup_{\alpha}\!L_\alpha$, $W=\bigcup_{\alpha}\! W_\alpha$,
where
$L_0=W_0=\emptyset$ and
\vspace{-.5ex}
\[
\begin{array}{l}
L_{\alpha+1}=
\{\, u \mid \forall_v\, mo v(u,v)  \Rightarrow v\in W_{\alpha}\,\},
\\
L_\beta=\bigcup_{\alpha<\beta}L_\alpha\ \  \mbox{for a limit ordinal } \beta,
\\
  W_\alpha= \{\, u \mid \exists_v\
       mo v(u,v)  \land v\in L_{\alpha}\,\}.
\end{array}
\smallskip
\]
(For game graphs of finite degree, i.e.\ with finite set of possible moves
from any position, standard sequences are sufficient.)
For finite ordinals, 
$L_i$ is the set of the losing positions with a failure in $\leq 2i-2$ moves,
and 
$W_i$ of the winning positions with a win in $\leq 2i-1$ moves.

Assume that $mo v$/2 is given by a finite set of facts.
(So the sequences are constant after a finite number of steps:
for some $j\in\NN$ and any $\alpha\mathop>j$,\, $L_\alpha=L_j=L$ and $W_\alpha=W_j=W$.
Then under the Prolog selection rule and SLS-resolution or SLDNF-resolution,
 ${\it WIN}$ does not diverge on atoms from $\st\cup (\HU\setminus\snf)$.
The proof is by induction on $i$, using 
the fact that for each such atom $w(u)$, $u\in W_i\cup L_i$ for some $i>0$.
Hence 
${\it WIN}$ is complete w.r.t.\ $spec$ under WFS.
Similarly, ${\it WIN}$ is complete w.r.t.\ $spec$ under KS for ground queries.

Hence
${\it WIN}$ is complete for the atoms from $\st\cup (\HU\setminus\snf)$
 w.r.t.\ $spec$ under WFS and KS.
(Note that swapping the two sets of $spec$ results in a proper
specification  $(\st,\snf)$, and 
${\it WIN}$ is complete and correct w.r.t.\ $(\st,\snf)$.)

\end{example}

\end{document}

%% file: diagram-approximate-neg.tex
%
%
\begin{minipage}{.7\textwidth}
$  
%
\newlength{\unit}\newlength{\vunit}  
 \setlength{\vunit}{.7cm}
 \setlength{\unit}{1cm}
\begin{array}{l}
\hspace{1pt}
     \overbrace{\rule{5.9\unit}{0pt}}^
         {\raisebox{.5ex}{\makebox[0pt][c]
             {\footnotesize
                   \begin{oldtabular}{c}
                      may succeed \\ \st
                   \end{oldtabular}%
           }}%
         }
     \\ [-.5ex]
\hspace{1pt}%
\overbrace{\rule{2.92\unit}{0pt}}^
    {\raisebox{.5ex}{\makebox[0pt][c]
        {\footnotesize \snf}}
    }%
\hspace{.02\unit}%
\grey
\aoverbrace[L2U1R]{\rule{5.9\unit}{0pt}}^
    {\raisebox{.5ex}{\makebox[0pt][c]
        {\footnotesize\black may fail}}
    }%
\\ [-1ex]
\framebox[3\unit]{\parbox[t][2\vunit][c]{3\unit}{\hfil
    \begin{oldtabular}{c}
       should succeed \\ \footnotesize cannot fail
    \end{oldtabular}%
    \hfil%
}}
\hspace{-.5pt}
\rlap{\mypalegrey\vrule height.15\vunit depth2.15\vunit width2.98\unit}%
\rlap{\hspace{.68\unit}\raisebox{-2.18\vunit}[0pt][0pt]{%
                     \makebox[0pt][c]{%
                  \includegraphics[trim=4.8cm 24cm 16.4cm 2.1cm,clip,scale=.96]
                   {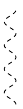}%
                   }}
}
\rlap{\hspace{2.32\unit}\raisebox{-2.18\vunit}[0pt][0pt]{%
                     \makebox[0pt][c]{%
                  \includegraphics[trim=4.8cm 24cm 16.4cm 2.1cm,clip,scale=.96]
                   {fala}%
                   }}
}
\framebox[3\unit]{\parbox[t][2\vunit][c]{3\unit}
    { \hfil \raisebox{.5ex}{irrelevant}\hfil }%
}  
\hspace{-.5pt}
%
\framebox[3\unit]{\parbox[t][2\vunit][c]{3\unit}{\hfil 
    \begin{oldtabular}{c}
      should fail \\ \footnotesize cannot succeed
    \end{oldtabular}%
    \hfil}}
 \makebox[0pt][l]{%
     \raisebox{-1.15\vunit}
              {\ $\left.\rule{0pt}{1.3\vunit}\right\}\HB$}%
     }
\\ [-1ex]
\hspace{1pt}%
\grey
\rlap{$\underbrace{\rule{3.65\unit}{0pt}}_
     {\raisebox{-.5ex}{\makebox[0pt][c]
         {\footnotesize\black $P$ succeeds}}%
     }
$}%
\hspace{5.32\unit}\hspace{-.12\unit}%
     \underbrace{\rule{3.68\unit}{0pt}}_
         {\raisebox{.5ex}{\makebox[0pt][c]
             {\footnotesize\black $P$ fails}}
         }
\end{array}
$    
\end{minipage}%
%